\documentclass[sigconf]{acmart}
\usepackage{wrapfig}

\AtBeginDocument{%
  \providecommand\BibTeX{{%
    \normalfont B\kern-0.5em{\scshape i\kern-0.25em b}\kern-0.8em\TeX}}}

\copyrightyear{2022}
\acmYear{2022}
\setcopyright{rightsretained}
\acmConference[TEI '22]{Sixteenth International Conference on Tangible, Embedded, and Embodied Interaction}{February 13--16, 2022}{Daejeon, Republic of Korea}
\acmBooktitle{Sixteenth International Conference on Tangible, Embedded, and Embodied Interaction (TEI '22), February 13--16, 2022, Daejeon, Republic of Korea}
\acmDOI{10.1145/3490149.3502424}
\acmISBN{978-1-4503-9147-4/22/02}

\begin{document}

\title{Redycler: Daily Outfit Texture Fabrication Appliance Using Re-Programmable Dyes}

\author{Ritik Batra}
\orcid{0000-0002-4476-5107}
\email{ritikbatra@berkeley.edu}
\authornotemark[1]
\author{Kaitlyn Lee}
\orcid{0000-0002-7869-6613}
\authornote{Both authors contributed equally to this research.}
\email{klee376@berkeley.edu}
\affiliation{%
  \institution{University of California, Berkeley}
  \city{Berkeley}
  \state{California}
  \country{USA}
  \postcode{94720}
}

\begin{abstract}
    We present a speculative design for a novel appliance for future fabrication in the home to revitalize textiles using re-programmable multi-color textures. Utilizing colored photochromic dyes activated by ultraviolet (UV) light, we can selectively deactivate hues using complementary colors in visible light to result in the final desired dye pattern. Our proposed appliance would automate this process within a box placed in the bedroom. We envision a future where people are able to transform old apparel into unique and fashionable pieces of clothing.
    
    We discuss how the user would interact with the appliance and how this device elongates the life-cycle of clothing through modification. We also outline the central issues to integrate such a concept into the home. Finally, we analyze how this device fits into personal modification trends in HCI to show how this device could change existing conceptions around sustainable fashion and personal style.
\end{abstract}



\maketitle

\section{Introduction}

There are several sustainability issues in the current state of the fashion industry.
It is estimated that the industry not only produces 8-10\% of global CO\textsubscript{2} emissions but is also a major consumer of water (79 trillion litres per year). It also produces more than 92 million tons of textile waste each year \cite{textileWaste}. In addition to the waste produced during garment production, clothing discarded after that process also contribute to a large amount of textile waste. The lives of several garment types (T-shirts, knit collared shirts, and woven pants) averaged only 3.1 to 3.5 years per garment, and only 15\% of textile waste is explicitly recycled globally \cite{textileWaste}.

 In today's highly competitive fashion market, retailers aim to keep up with the latest trends in fashion shows, leading to them increasing the frequency with which they rotate out their merchandise in the destructive business model today known as \emph{fast fashion} \cite{doi:10.1080/09593960903498300}. This encourages customers to buy clothing more frequently with the idea of "Here today, gone tomorrow."
 
 Many existing fashion brands, like Zara and H\&M pride themselves on keeping up to date with the hottest trends, boasting impressively short production cycles. Shein, a rapidly growing online fast fashion brand, aims to get products from design to shipment within 3 days \cite{fastfashion}.

\begin{figure}
    \begin{center}
        \includegraphics[width=.5\textwidth]{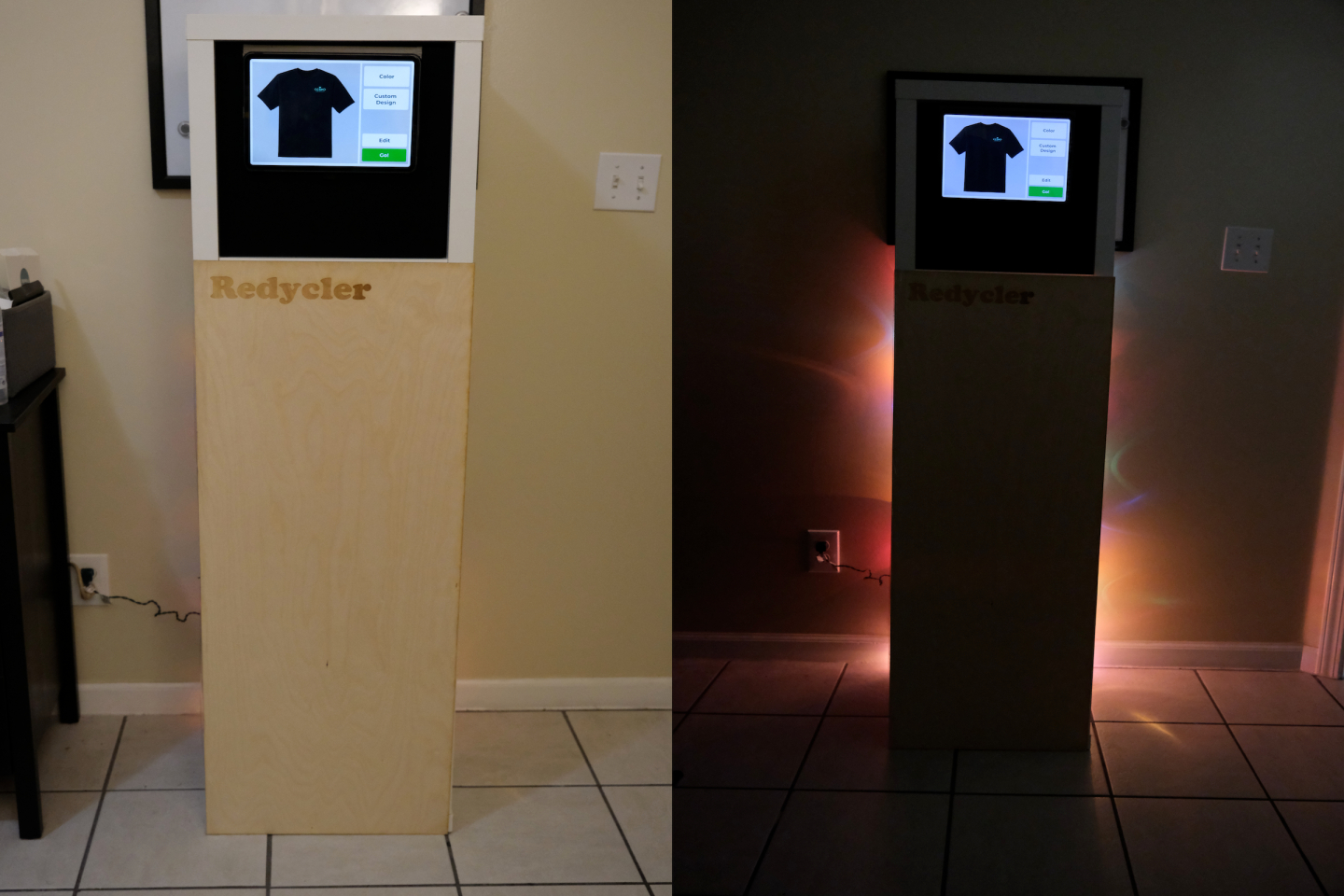}
        \caption{This showcases our speculative appliance design placed in a bedroom.}
        \label{high_res}
    \end{center}
\end{figure}

Hence, we propose a speculative design shown in Figure \ref{high_res} for a new common household appliance that allows users to quickly and easily change the color and designs of articles of clothing. We aim to show how this appliance integrates into everyday life and show how it is easy to use by incorporating familiar interactions. 

This device challenges the sustainability issues of fast fashion by allowing users to change styles and revitalize old pieces of clothing without needing to create new textiles, while empowering users to express themselves through personal styling \cite{alteration}.

Focusing on personal fabrication, this appliance would enable people outside of technology enthusiast communities to modify their clothing without requiring extensive setup or skill acquisition \cite{ubiquitous}. Through a simple interface, we hope to lower the barrier for personal fabrication and encourage more tailored textiles rather than wastefully purchasing new ones.

Our research questions are as follows. 

\begin{enumerate}
  \item How do we encourage self-expression in clothing through personal unique modifications?
  \item How can we motivate consumers to be more sustainable and recycle their clothing?
  \item What interactions can we use to fully integrate this novel machine into regular household routines?
\end{enumerate}

Of course, there are also consumers for whom staying completely up to date with the latest trends is not a priority. For such people, thrift shopping is a popular way of refreshing a wardrobe. A recent study on young thrift-shoppers in Rio De Janeiro found that people thrift shopped for a mixture of three reasons--price, exclusivity, and quality~\cite{thrifting}. However, thrift shopping is also quite restrictive; shoppers may not be able to find a design they like in the size they need. Here, we propose a device that takes a garment a person already owns as input -- perhaps a faded but comfortable sweater -- and allows the owner to programmatically re-color the garment in their own home. Such a device would reduce the frustration of finding clothing that is simultaneously form-fitting, stylistically refreshing, and eco-friendly all at once.

\section{Related Work}
Qamar et al. have demonstrated a method of using photochromic dyes to programmatically color and re-color single-material objects with Photo-Chromeleon. Their work introduces the reversible coloration of textiles using photochromic dye, ultraviolet (UV) light, and visible light but we would like to integrate this technology into the daily life of users using an automated, interactive appliance that can be used daily \cite{PhotoChromeleon}. The ultra-short throw projector with a UV LED wash in the box in contrast to a UV and visible light projector set a specific distance away would fit better into the home due to space required and workflow automation. We are also lowering the barrier of entry for users with a drag-and-drop user interface and 2D design uploads in contrast to the Photo-Chromeleon 3D UV mappings.

This design builds on previous work by Devendorf et al. relating to dynamic textiles, as we are connecting textile reconstruction with personal style and exploring how users would interact with an appliance enabling them to automatically make this transformation, as well as how dynamic textiles can be integrated into daily life \cite{perception}. 

A challenge we see with this integration is the motivation for users to spend the extra time to use this device rather than resorting to going to a retailer and purchasing a new garment, so we hope to eventually incentivize users through a rewards-based system similar to how we motivate people to purchase electric cars and renewable sources of energy through tax breaks \cite{playful}. By using gamification and technology to persuade users to act more sustainably, we introduce a playful interface and fun user experience that motivates users to regularly utilize the device \cite{gamification}. To further motivate the user to recycle clothing, the appliance's interface includes metrics of the positive environmental impact by using this appliance such as tons of textile waste prevented.

\section{Technology}

\subsection{Hardware}


A photochromic dye becomes saturated ("activated") when exposed to UV light, so the combined cyan, magenta, and yellow (CMY) dye first becomes black. Since each CMY component has a unique absorption peak in the visible spectrum, the dyes can then be selectively and rapidly de-saturated ("deactivated") with close-range red, green, and/or blue (RGB) visible light to result in a desired color \cite{PhotoChromeleon}. By combining CMY photochromic dyes, we can attach a spray to the device that can be applied to the entire garment by the user before operating the appliance. With close-range projectors, we can improve the speed and integration of re-programming the dye for daily personal fabrication.

\begin{figure}
    \begin{center}
        \includegraphics[width=0.3\textwidth]{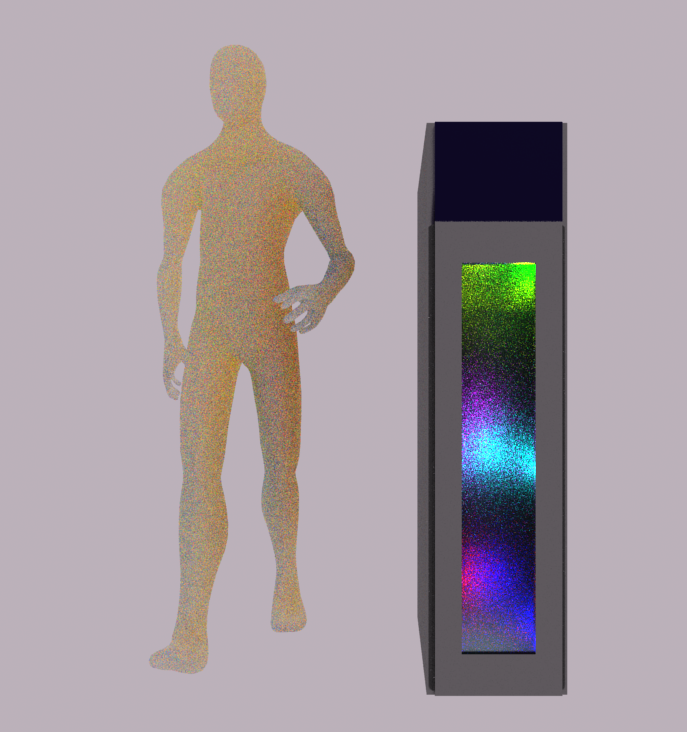}
        \caption{We rendered a 3D model to visualize how this device would look like next to a humanoid.}
        \label{maya}
    \end{center}
\end{figure}

A mock-up of the appliance is shown in Figure \ref{maya}. Redycler consists of a large sealed door and an easy-to-use digital interface to quickly select colors and patterns to apply to the chosen garment. The technology would consist of a touchscreen compatible with a Raspberry Pi to provide a playful graphical user interface to provide the service to users in an engaging way.

To fully integrate this appliance into daily life, setting up the material for dyeing should require minimal user effort. The user would need to initially spray the garments with photochromic dye which would be included in a refillable spray bottle attached to the appliance. There would be two clips to hang the piece on one side of the interior and an ultra short throw projector below the piece to project the visible light onto the garment. The clips would ensure that there are no folds and creases in the print. The interior would first be flooded with UV light from LEDs to activate the CMY dyes and turn the piece black. Then, the projector selectively projects RGB visible light to deactivate specific colors of the piece until the desired design is created. Since we are using a projector, we are able to maintain the resolution of our design to be at least 720p. We would also fit a camera into the interior of the appliance on the opposing wall of the garment to approximate the amount of material to dye based on garment size and design complexity which allows the device to estimate the time to completion and provide a preview to the user on the interface \cite{image_size}. With visuals of the garment, we can best approximate what dye added to the existing color of the piece would produce the desired hue though color mixing algorithms.

In regards to appliance maintenance, you would only need to spray the garment with the dyes to refresh it every few washes. Washing the garment inside-out would not remove the dye if given enough time for the dye to settle into the garment.

\subsection{User Interface}

\begin{figure}
    \begin{center}
        \includegraphics[width=0.5\textwidth]{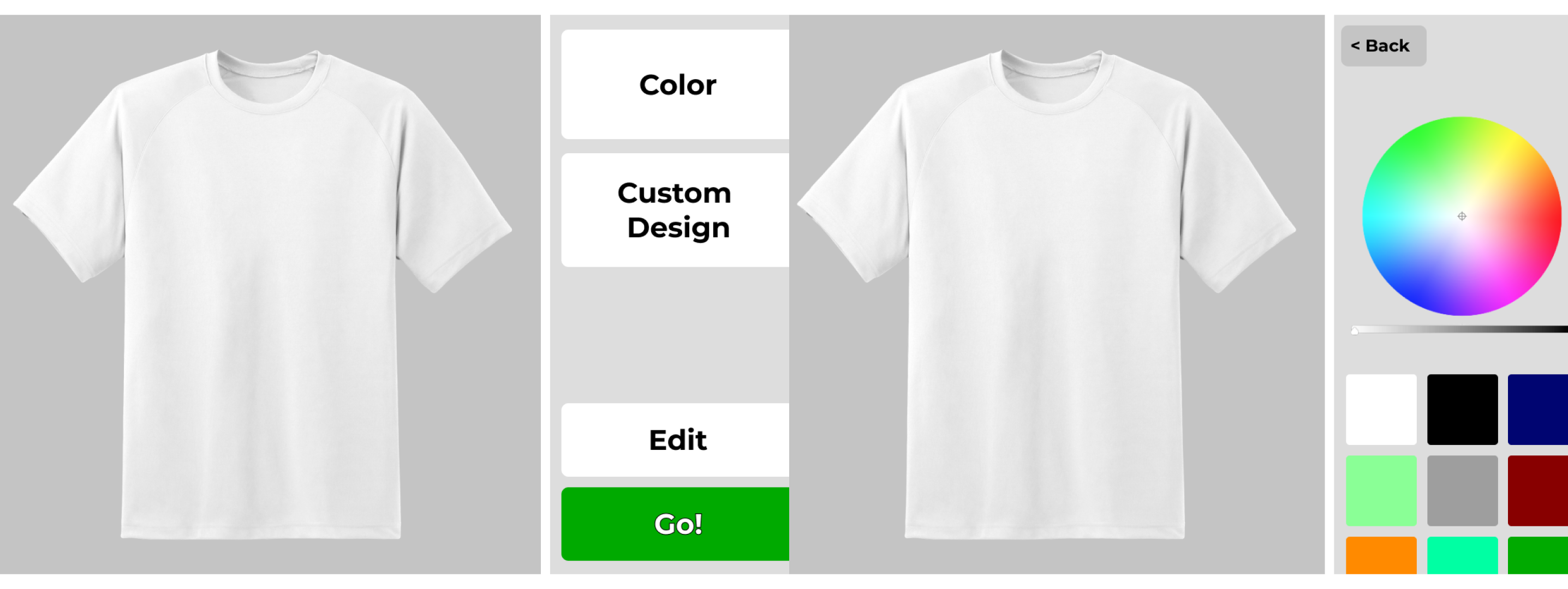}
        \caption{Users can see what the garment looks like and preview select colors and designs.}
        \label{figma}
    \end{center}
\end{figure}

We also propose a web interface to upload new designs and share with the Redycler community. The simple user interface in Figure \ref{figma} shows the user what their current design looks like and allows them to pick any color to dye their selected garment. The interface on the top would prompt the user to make selections of what design to apply as well as the time remaining during a print. In addition, the user can also choose to add a custom design onto their apparel. These custom designs can also be uploaded and shared through an online platform. In order to expand personal fabrication to a wider range of users, it is more accessible to simplify the process of selecting a finished product, rather than trying to simplify the modeling process \cite{9313032}. As such, we chose to have users upload their designs and share them with others in order to make obtaining a design as easy as possible and build a community among users. This will also connect people of various backgrounds together and inspire new fashion patterns and styles. 

\section{Limitations}
In terms of technical limitations today, we are restricted by the durability of these dyes' activation because daily routines usually involve natural sunlight which may activate or deactivate the shirt accordingly. We also would explore increasing the speed of this fabrication device to reach time requirements of users who want to change their clothes in the morning. There is work in this space with gray-scale previews being printed in less than a minute by selectively saturating the dye with a UV projector, so potential future work could be integrating rapid prototyping into our speculative design \cite{chromoupdate}. We also have to consider the limitation of some colors having to be approximated due to the existing color or dye of the garment.

\section{Scenarios}

We imagined several scenarios for the use of this at-home appliance for re-programming the dyes of a garment, such as customization to revitalize a garment that the user has lost interest in or has gone "out of style", quickly shifting the style of a garment between day and night use, and modifying garments to emphasize the user's personal sense of style. 

As a concrete example, a user can select a t-shirt to modify, place it in the appliance, and then select a design that they desire for the garment. After a few minutes of fabrication, the user can retrieve their garment and wear it throughout the day. At the end of the day, the user can reset the garment and select a new design to print. This cycle of re-printing on the garment is what promotes sustainable fashion through personal modification of the same material.

\begin{figure*}
    \begin{center}
         \includegraphics[width=\textwidth]{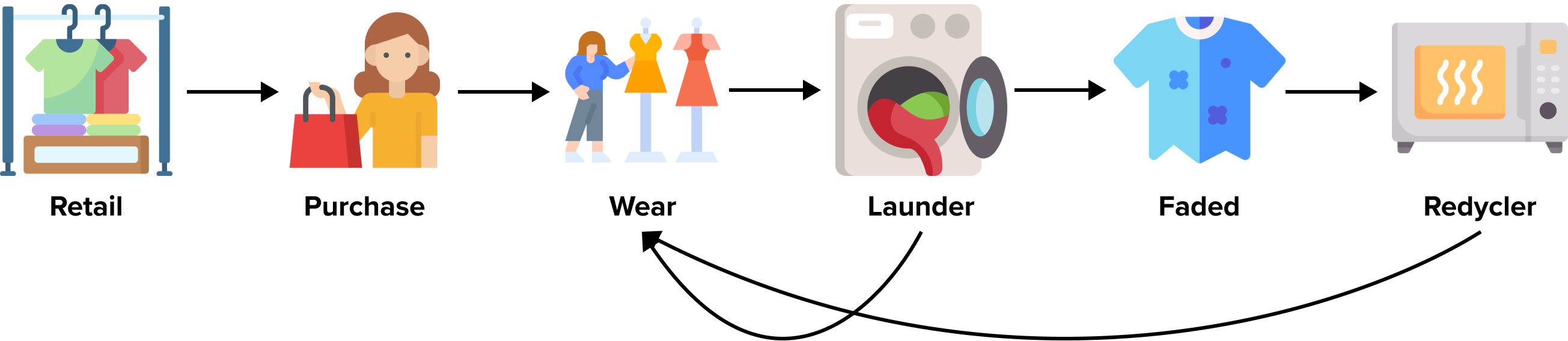}
        \caption{This is the re-imagined life-cycle of an article of clothing with this speculative design.}
        \label{user_flow}
    \end{center}
\end{figure*}

One of the goals of this design is that the device could be easily incorporated into the user's daily routine. The device is the size of a small bookshelf, meaning that it is easy to place in a closet or bedroom. Figure \ref{user_flow} details the journey of a garment with the integration of this novel appliance.

\section{Discussion}
Integrating the personal programmable modification of apparel into daily life would shift the way we buy clothing because the focus of consumerism would be on quality over clothing style. Fast fashion today is a cheap and unsustainable way for businesses to keep up with ever-changing fashion trends, but this design would challenge that industry direction.

We hope this design will also promote self-expression through the personal modification of outfits as clothing and identity have always been intertwined \cite{selfexpression}. Enabling people to be able to customize their clothing based on their emotions is another direction fashion could take with this design as there are links between fashion purchasing process and emotions \cite{emotion}.

This speculative design encourages a new community of creators to personalize the material goods they are consuming and share those designs with others. By creating a place for users to share their unique designs, we hope to build a community that cyclically inspires each other to create personalized outfits and shifts the reliance on creativity from textile manufacturers to individual creators \cite{democratization}.

\section{Conclusion and Future Work}
Textile waste is growing in intensity as landfills expand and society becomes increasingly consumerist and we would like to challenge this through personal fabrication. Normally, if a garment becomes old or boring, it goes to either the back of the closet or the garbage. We propose an alternative home appliance that is placed in the bedroom next to a user's closet. This appliance enables users to refresh their clothing with customized colors and designs, promoting self-expression through evolving style while prolonging a garment's lifespan. The revitalization process is possible through photochromic dye and the selective activation and deactivation of colors to set the color of the garment with high resolution and precision.

Similar to the recent personal computing boom, we are on the cusp of personal fabrication becoming increasingly integrated into the home \cite{personal_fabrication}. Customization of tailored artifacts is just one of the many incentives for pushing personal fabrication towards modifications \cite{ubiquitous}. We envision a new garment life-cycle that is more sustainable and enables users to reuse the same fabric with constantly changing appearances. This customization, with performance and technical enhancements, would spark more work in fashion sustainability and focus attention on garment comfort and shape rather than branding and color.

Future designs would shift beyond just changing textile colors and patterns to the shape of clothing to enable users to change how their clothing fits and what aligns with current fashion style trends.

This trend also encourages future work in dynamically-changing and reactive fashion to connect color theory to moods in older people or even last-minute modifications to outfits and colors based on event dress codes \cite{colortheory}. We hope that our appliance design will inspire further development in the sustainable modification and personalization of clothing to adapt to a user's needs and style. 

\begin{acks}
To Sarah Sterman, Katherine Song, Eric Rawn, Molly Nicholas, and Professor Eric Paulos, for guiding and mentoring us through this exciting process.
\end{acks}

\bibliographystyle{ACM-Reference-Format}
\bibliography{main}


\begin{thebibliography}{18}


\ifx \showCODEN    \undefined \def \showCODEN     #1{\unskip}     \fi
\ifx \showDOI      \undefined \def \showDOI       #1{#1}\fi
\ifx \showISBNx    \undefined \def \showISBNx     #1{\unskip}     \fi
\ifx \showISBNxiii \undefined \def \showISBNxiii  #1{\unskip}     \fi
\ifx \showISSN     \undefined \def \showISSN      #1{\unskip}     \fi
\ifx \showLCCN     \undefined \def \showLCCN      #1{\unskip}     \fi
\ifx \shownote     \undefined \def \shownote      #1{#1}          \fi
\ifx \showarticletitle \undefined \def \showarticletitle #1{#1}   \fi
\ifx \showURL      \undefined \def \showURL       {\relax}        \fi
\providecommand\bibfield[2]{#2}
\providecommand\bibinfo[2]{#2}
\providecommand\natexlab[1]{#1}
\providecommand\showeprint[2][]{arXiv:#2}

\bibitem[\protect\citeauthoryear{Baudisch and Mueller}{Baudisch and
  Mueller}{2017}]%
        {personal_fabrication}
\bibfield{author}{\bibinfo{person}{Patrick Baudisch} {and}
  \bibinfo{person}{Stefanie Mueller}.} \bibinfo{year}{2017}\natexlab{}.
\newblock \showarticletitle{Personal Fabrication}.
\newblock \bibinfo{journal}{\emph{Foundations and Trends® in Human–Computer
  Interaction}} \bibinfo{volume}{10}, \bibinfo{number}{3–4}
  (\bibinfo{year}{2017}), \bibinfo{pages}{165--293}.
\newblock
\showISSN{1551-3955}
\urldef\tempurl%
\url{https://doi.org/10.1561/1100000055}
\showDOI{\tempurl}


\bibitem[\protect\citeauthoryear{Bhardwaj and Fairhurst}{Bhardwaj and
  Fairhurst}{2010}]%
        {doi:10.1080/09593960903498300}
\bibfield{author}{\bibinfo{person}{Vertica Bhardwaj} {and} \bibinfo{person}{Ann
  Fairhurst}.} \bibinfo{year}{2010}\natexlab{}.
\newblock \showarticletitle{Fast fashion: response to changes in the fashion
  industry}.
\newblock \bibinfo{journal}{\emph{The International Review of Retail,
  Distribution and Consumer Research}} \bibinfo{volume}{20},
  \bibinfo{number}{1} (\bibinfo{year}{2010}), \bibinfo{pages}{165--173}.
\newblock
\urldef\tempurl%
\url{https://doi.org/10.1080/09593960903498300}
\showDOI{\tempurl}
\showeprint{https://doi.org/10.1080/09593960903498300}


\bibitem[\protect\citeauthoryear{Bosch and Kanis}{Bosch and Kanis}{2013}]%
        {playful}
\bibfield{author}{\bibinfo{person}{Lilian Bosch} {and} \bibinfo{person}{Marije
  Kanis}.} \bibinfo{year}{2013}\natexlab{}.
\newblock \showarticletitle{Encouraging Sustainable Fashion with a Playful
  Recycling System}. In \bibinfo{booktitle}{\emph{Proceedings of the 27th
  International BCS Human Computer Interaction Conference}} (London, UK)
  \emph{(\bibinfo{series}{BCS-HCI '13})}. \bibinfo{publisher}{BCS Learning \&
  Development Ltd.}, \bibinfo{address}{Swindon, GBR}, Article
  \bibinfo{articleno}{46}, \bibinfo{numpages}{6}~pages.
\newblock


\bibitem[\protect\citeauthoryear{Cardoso, Ribeiro, Prandi, and Nunes}{Cardoso
  et~al\mbox{.}}{2019}]%
        {gamification}
\bibfield{author}{\bibinfo{person}{Bruno Cardoso}, \bibinfo{person}{Miguel
  Ribeiro}, \bibinfo{person}{Catia Prandi}, {and} \bibinfo{person}{Nuno
  Nunes}.} \bibinfo{year}{2019}\natexlab{}.
\newblock \showarticletitle{When Gamification Meets Sustainability: A Pervasive
  Approach to Foster Sustainable Mobility in Madeira}. In
  \bibinfo{booktitle}{\emph{Proceedings of the 1st ACM Workshop on Emerging
  Smart Technologies and Infrastructures for Smart Mobility and
  Sustainability}} (Los Cabos, Mexico) \emph{(\bibinfo{series}{SMAS '19})}.
  \bibinfo{publisher}{Association for Computing Machinery},
  \bibinfo{address}{New York, NY, USA}, \bibinfo{pages}{3–8}.
\newblock
\showISBNx{9781450369305}
\urldef\tempurl%
\url{https://doi.org/10.1145/3349622.3355449}
\showDOI{\tempurl}


\bibitem[\protect\citeauthoryear{Cegindir and Bayram}{Cegindir and
  Bayram}{2018}]%
        {alteration}
\bibfield{author}{\bibinfo{person}{Nese Cegindir} {and} \bibinfo{person}{Ankara
  Bayram}.} \bibinfo{year}{2018}\natexlab{}.
\newblock \showarticletitle{An analysis of alteration in design from garment to
  fashion product}.
\newblock


\bibitem[\protect\citeauthoryear{Corrêa and Dubeux}{Corrêa and
  Dubeux}{2015}]%
        {thrifting}
\bibfield{author}{\bibinfo{person}{Sílvia Corrêa} {and}
  \bibinfo{person}{Veranise Dubeux}.} \bibinfo{year}{2015}\natexlab{}.
\newblock \showarticletitle{Buying clothes from thrift stores: an analysis of
  young people consuming second-hand clothing in Rio de Janeiro}.
\newblock \bibinfo{journal}{\emph{Comunicação, Mídia e Consumo}}
  \bibinfo{volume}{12} (\bibinfo{date}{05} \bibinfo{year}{2015}).
\newblock
\urldef\tempurl%
\url{https://doi.org/10.18568/1983-7070.123334-56}
\showDOI{\tempurl}


\bibitem[\protect\citeauthoryear{Devendorf, Lo, Howell, Lee, Gong, Karagozler,
  Fukuhara, Poupyrev, Paulos, and Ryokai}{Devendorf et~al\mbox{.}}{2016}]%
        {perception}
\bibfield{author}{\bibinfo{person}{Laura Devendorf}, \bibinfo{person}{Joanne
  Lo}, \bibinfo{person}{Noura Howell}, \bibinfo{person}{Jung~Lin Lee},
  \bibinfo{person}{Nan-Wei Gong}, \bibinfo{person}{M.~Emre Karagozler},
  \bibinfo{person}{Shiho Fukuhara}, \bibinfo{person}{Ivan Poupyrev},
  \bibinfo{person}{Eric Paulos}, {and} \bibinfo{person}{Kimiko Ryokai}.}
  \bibinfo{year}{2016}\natexlab{}.
\newblock \showarticletitle{"I Don't Want to Wear a Screen": Probing
  Perceptions of and Possibilities for Dynamic Displays on Clothing}. In
  \bibinfo{booktitle}{\emph{Proceedings of the 2016 CHI Conference on Human
  Factors in Computing Systems}} (San Jose, California, USA)
  \emph{(\bibinfo{series}{CHI '16})}. \bibinfo{publisher}{Association for
  Computing Machinery}, \bibinfo{address}{New York, NY, USA},
  \bibinfo{pages}{6028–6039}.
\newblock
\showISBNx{9781450333627}
\urldef\tempurl%
\url{https://doi.org/10.1145/2858036.2858192}
\showDOI{\tempurl}


\bibitem[\protect\citeauthoryear{Giraldo, Casta\~{n}o, Giraldo, and
  Mej\'{\i}a}{Giraldo et~al\mbox{.}}{2019}]%
        {colortheory}
\bibfield{author}{\bibinfo{person}{F\'{a}ber~D. Giraldo},
  \bibinfo{person}{Esteban~M. Casta\~{n}o}, \bibinfo{person}{Sebasti\'{a}n
  Giraldo}, {and} \bibinfo{person}{Sebasti\'{a}n Mej\'{\i}a}.}
  \bibinfo{year}{2019}\natexlab{}.
\newblock \showarticletitle{Literature Review on the Theory of Color and Its
  Relationship with Moods in Older People}. In
  \bibinfo{booktitle}{\emph{Proceedings of the 5th Workshop on ICTs for
  Improving Patients Rehabilitation Research Techniques}} (Popayan, Columbia)
  \emph{(\bibinfo{series}{REHAB '19})}. \bibinfo{publisher}{Association for
  Computing Machinery}, \bibinfo{address}{New York, NY, USA},
  \bibinfo{pages}{15–18}.
\newblock
\showISBNx{9781450371513}
\urldef\tempurl%
\url{https://doi.org/10.1145/3364138.3364144}
\showDOI{\tempurl}


\bibitem[\protect\citeauthoryear{Insights}{Insights}{2021}]%
        {fastfashion}
\bibfield{author}{\bibinfo{person}{CB Insights}.}
  \bibinfo{year}{2021}\natexlab{}.
\newblock \bibinfo{booktitle}{\emph{The Future Of Fashion: From Design To
  Merchandising, How Tech Is Reshaping The Industry}}.
\newblock
\urldef\tempurl%
\url{https://www.cbinsights.com/research/fashion-tech-future-trends}
\showURL{%
\tempurl}


\bibitem[\protect\citeauthoryear{Jin, Qamar, Wessely, Adhikari, Bulovic,
  Punpongsanon, and Mueller}{Jin et~al\mbox{.}}{2019}]%
        {PhotoChromeleon}
\bibfield{author}{\bibinfo{person}{Yuhua Jin}, \bibinfo{person}{Isabel Qamar},
  \bibinfo{person}{Michael Wessely}, \bibinfo{person}{Aradhana Adhikari},
  \bibinfo{person}{Katarina Bulovic}, \bibinfo{person}{Parinya Punpongsanon},
  {and} \bibinfo{person}{Stefanie Mueller}.} \bibinfo{year}{2019}\natexlab{}.
\newblock \showarticletitle{Photo-Chromeleon: Re-Programmable Multi-Color
  Textures Using Photochromic Dyes}. In \bibinfo{booktitle}{\emph{Proceedings
  of the 32nd Annual ACM Symposium on User Interface Software and Technology}}
  (New Orleans, LA, USA) \emph{(\bibinfo{series}{UIST '19})}.
  \bibinfo{publisher}{Association for Computing Machinery},
  \bibinfo{address}{New York, NY, USA}, \bibinfo{pages}{701–712}.
\newblock
\showISBNx{9781450368162}
\urldef\tempurl%
\url{https://doi.org/10.1145/3332165.3347905}
\showDOI{\tempurl}


\bibitem[\protect\citeauthoryear{Li, Xu, Xiao, Liu, Feng, and Zhang}{Li
  et~al\mbox{.}}{2017}]%
        {image_size}
\bibfield{author}{\bibinfo{person}{Chunxiao Li}, \bibinfo{person}{Ying Xu},
  \bibinfo{person}{Yi Xiao}, \bibinfo{person}{Huimin Liu},
  \bibinfo{person}{Meiling Feng}, {and} \bibinfo{person}{Dongliang Zhang}.}
  \bibinfo{year}{2017}\natexlab{}.
\newblock \showarticletitle{Automatic Measurement of Garment Sizes Using Image
  Recognition}. In \bibinfo{booktitle}{\emph{Proceedings of the International
  Conference on Graphics and Signal Processing}} (Singapore, Singapore)
  \emph{(\bibinfo{series}{ICGSP '17})}. \bibinfo{publisher}{Association for
  Computing Machinery}, \bibinfo{address}{New York, NY, USA},
  \bibinfo{pages}{30–34}.
\newblock
\showISBNx{9781450352390}
\urldef\tempurl%
\url{https://doi.org/10.1145/3121360.3121382}
\showDOI{\tempurl}


\bibitem[\protect\citeauthoryear{Mota}{Mota}{2011}]%
        {democratization}
\bibfield{author}{\bibinfo{person}{Catarina Mota}.}
  \bibinfo{year}{2011}\natexlab{}.
\newblock \showarticletitle{The Rise of Personal Fabrication}. In
  \bibinfo{booktitle}{\emph{Proceedings of the 8th ACM Conference on Creativity
  and Cognition}} (Atlanta, Georgia, USA) \emph{(\bibinfo{series}{C\&C '11})}.
  \bibinfo{publisher}{Association for Computing Machinery},
  \bibinfo{address}{New York, NY, USA}, \bibinfo{pages}{279–288}.
\newblock
\showISBNx{9781450308205}
\urldef\tempurl%
\url{https://doi.org/10.1145/2069618.2069665}
\showDOI{\tempurl}


\bibitem[\protect\citeauthoryear{Niinim{\"a}ki, Peters, Dahlbo, Perry,
  Rissanen, and Gwilt}{Niinim{\"a}ki et~al\mbox{.}}{2020}]%
        {textileWaste}
\bibfield{author}{\bibinfo{person}{Kirsi Niinim{\"a}ki}, \bibinfo{person}{Greg
  Peters}, \bibinfo{person}{Helena Dahlbo}, \bibinfo{person}{Patsy Perry},
  \bibinfo{person}{Timo Rissanen}, {and} \bibinfo{person}{Alison Gwilt}.}
  \bibinfo{year}{2020}\natexlab{}.
\newblock \showarticletitle{The environmental price of fast fashion}.
\newblock \bibinfo{journal}{\emph{Nature Reviews Earth \& Environment}}
  \bibinfo{volume}{1}, \bibinfo{number}{4} (\bibinfo{year}{2020}),
  \bibinfo{pages}{189--200}.
\newblock
\showISBNx{2662-138X}
\urldef\tempurl%
\url{https://doi.org/10.1038/s43017-020-0039-9}
\showDOI{\tempurl}


\bibitem[\protect\citeauthoryear{Piazza, Kr\"{o}ckel, and Bodendorf}{Piazza
  et~al\mbox{.}}{2017}]%
        {emotion}
\bibfield{author}{\bibinfo{person}{Alexander Piazza}, \bibinfo{person}{Pavlina
  Kr\"{o}ckel}, {and} \bibinfo{person}{Freimut Bodendorf}.}
  \bibinfo{year}{2017}\natexlab{}.
\newblock \showarticletitle{Emotions and Fashion Recommendations: Evaluating
  the Predictive Power of Affective Information for the Prediction of Fashion
  Product Preferences in Cold-Start Scenarios}. In
  \bibinfo{booktitle}{\emph{Proceedings of the International Conference on Web
  Intelligence}} (Leipzig, Germany) \emph{(\bibinfo{series}{WI '17})}.
  \bibinfo{publisher}{Association for Computing Machinery},
  \bibinfo{address}{New York, NY, USA}, \bibinfo{pages}{1234–1240}.
\newblock
\showISBNx{9781450349512}
\urldef\tempurl%
\url{https://doi.org/10.1145/3106426.3109441}
\showDOI{\tempurl}


\bibitem[\protect\citeauthoryear{Roinesalo, Rantakari, Virtanen, and
  H\"{a}kkil\"{a}}{Roinesalo et~al\mbox{.}}{2016}]%
        {selfexpression}
\bibfield{author}{\bibinfo{person}{Paula Roinesalo}, \bibinfo{person}{Juho
  Rantakari}, \bibinfo{person}{Lasse Virtanen}, {and} \bibinfo{person}{Jonna
  H\"{a}kkil\"{a}}.} \bibinfo{year}{2016}\natexlab{}.
\newblock \showarticletitle{Clothes Integrated Visual Markers as
  Self-Expression Tool}. In \bibinfo{booktitle}{\emph{Proceedings of the 18th
  International Conference on Human-Computer Interaction with Mobile Devices
  and Services Adjunct}} (Florence, Italy) \emph{(\bibinfo{series}{MobileHCI
  '16})}. \bibinfo{publisher}{Association for Computing Machinery},
  \bibinfo{address}{New York, NY, USA}, \bibinfo{pages}{617–620}.
\newblock
\showISBNx{9781450344135}
\urldef\tempurl%
\url{https://doi.org/10.1145/2957265.2961832}
\showDOI{\tempurl}


\bibitem[\protect\citeauthoryear{Stemasov, Rukzio, and Gugenheimer}{Stemasov
  et~al\mbox{.}}{2021a}]%
        {ubiquitous}
\bibfield{author}{\bibinfo{person}{Evgeny Stemasov}, \bibinfo{person}{Enrico
  Rukzio}, {and} \bibinfo{person}{Jan Gugenheimer}.}
  \bibinfo{year}{2021}\natexlab{a}.
\newblock \showarticletitle{The Road to Ubiquitous Personal Fabrication:
  Modeling-Free Instead of Increasingly Simple}.
\newblock \bibinfo{journal}{\emph{IEEE Pervasive Computing}}
  \bibinfo{volume}{20}, \bibinfo{number}{1} (\bibinfo{year}{2021}),
  \bibinfo{pages}{19--27}.
\newblock
\urldef\tempurl%
\url{https://doi.org/10.1109/MPRV.2020.3029650}
\showDOI{\tempurl}


\bibitem[\protect\citeauthoryear{Stemasov, Rukzio, and Gugenheimer}{Stemasov
  et~al\mbox{.}}{2021b}]%
        {9313032}
\bibfield{author}{\bibinfo{person}{E. Stemasov}, \bibinfo{person}{E. Rukzio},
  {and} \bibinfo{person}{J. Gugenheimer}.} \bibinfo{year}{2021}\natexlab{b}.
\newblock \showarticletitle{The Road to Ubiquitous Personal Fabrication:
  Modeling-Free Instead of Increasingly Simple}.
\newblock \bibinfo{journal}{\emph{IEEE Pervasive Computing}}
  \bibinfo{volume}{20}, \bibinfo{number}{01} (\bibinfo{date}{jan}
  \bibinfo{year}{2021}), \bibinfo{pages}{19--27}.
\newblock
\showISSN{1558-2590}
\urldef\tempurl%
\url{https://doi.org/10.1109/MPRV.2020.3029650}
\showDOI{\tempurl}


\bibitem[\protect\citeauthoryear{Wessely, Jin, Nuengsigkapian, Kashapov, Qamar,
  Tsetserukou, and Mueller}{Wessely et~al\mbox{.}}{2021}]%
        {chromoupdate}
\bibfield{author}{\bibinfo{person}{Michael Wessely}, \bibinfo{person}{Yuhua
  Jin}, \bibinfo{person}{Cattalyya Nuengsigkapian}, \bibinfo{person}{Aleksei
  Kashapov}, \bibinfo{person}{Isabel P.~S. Qamar}, \bibinfo{person}{Dzmitry
  Tsetserukou}, {and} \bibinfo{person}{Stefanie Mueller}.}
  \bibinfo{year}{2021}\natexlab{}.
\newblock \bibinfo{booktitle}{\emph{ChromoUpdate: Fast Design Iteration of
  Photochromic Color Textures Using Grayscale Previews and Local Color
  Updates}}.
\newblock \bibinfo{publisher}{Association for Computing Machinery},
  \bibinfo{address}{New York, NY, USA}.
\newblock
\showISBNx{9781450380966}
\urldef\tempurl%
\url{https://doi.org/10.1145/3411764.3445391}
\showURL{%
\tempurl}


\end{thebibliography}

\appendix

\end{document}